\documentclass[final]{l4dc2026}

\usepackage{times}
\usepackage{subcaption}
\usepackage{tikz}
\usepackage{amsmath}
\usepackage{url}
\usepackage{hyperref}
\usepackage{amssymb}
\usetikzlibrary{patterns}
\usepackage{algorithm2e}
\usetikzlibrary{positioning,shapes,arrows,calc}

\title[Learning to Control Misinformation]{Learning to Control Misinformation: a Closed-loop Approach for Misinformation Mitigation over Social Networks}

\author{%
 \Name{Nicolò Pagan} \Email{nicolo.pagan@uzh.ch}\\
 \addr Institut für Informatik\\
 \addr University of Zurich\\
 \addr Zurich, CH
 \AND
 \Name{Andreas Philippou}
  \Email{a.philippou@student.tue.nl}\\
 \addr Department of Electrical Engineering\\
 \addr Eindhoven University of Technology\\
 \addr Eindhoven, The Netherlands
 \AND
 \Name{Giulia De~Pasquale}
   \Email{g.de.pasquale@tue.nl}\\
 \addr Department of Electrical Engineering\\
 \addr Eindhoven University of Technology\\
 \addr Eindhoven, The Netherlands
}

\begin{document}
\maketitle

\begin{abstract}
Modern social networks rely on recommender systems that inadvertently amplify misinformation by prioritizing engagement over content veracity. We present a control framework that mitigates misinformation spread while maintaining user engagement by penalizing content characteristics commonly exploited by false information—specifically, extreme negative sentiment and novelty. 
We extend the closed-loop Friedkin-Johnsen model to incorporate the mitigation of misinformation  together with the maximization of user engagement.
Both model-free and model-based control strategies demonstrate up to 76\% reduction in misinformation propagation across diverse network configurations, validated through simulations using the LIAR2 dataset with sentiment features extracted via large language models. Analysis of engagement-misinformation trade-offs reveals that in networks with radical users, median engagement improves even as misinformation decreases, suggesting content moderation enhances discourse quality for non-extremist users. The framework provides practical guidance for platform operators 
in balancing misinformation suppression with engagement objectives.
\end{abstract}


\section{Introduction}

Modern social networks connect billions of users but simultaneously create conditions that facilitate rapid misinformation spread \citep{del2016spreading}. The societal consequences—affecting democratic processes, public health, and social cohesion \citep{cinelli2020covid, persily2017can}—have intensified as research confirms misinformation spreads faster than truth \citep{vosoughi2018spread}.

Current mitigation strategies focus on content truthfulness through fact-checking and machine learning \citep{shu2017fake}, yet overlook the psychological mechanisms driving viral spread. Misinformation strategically exploits emotional triggers—particularly negative emotions \citep{brady2017emotion}—and novelty \citep{berger2012makes} to achieve virality. Recommender systems designed to maximize engagement can amplify such content, creating feedback loops that reinforce echo chambers and polarization \citep{vicario2015echo, mansoury2020feedback, pagan2023classification, lanzetti2023impact}.

Recent advances model recommender systems as control inputs within opinion dynamics frameworks \cite{dean2024accounting,closed_loop_opinion,network_aware_rec_sys_via_feedback,dean2022preference,mansoury2020feedback,sprenger2024control}. In particular, \citet{sprenger2024control} developed a closed-loop Friedkin-Johnsen model where engagement-maximizing recommendations fundamentally alter network opinion evolution. However, research explicitly addressing misinformation within this control framework remains limited.

We adapt this framework to model \emph{sentiment propagation} rather than topical opinion, motivated by evidence that misinformation spreads through emotional manipulation. We modify the engagement objective to penalize extreme negative sentiment and novelty—characteristics misinformation exploits for virality—while maintaining user engagement. Both model-free and model-based strategies are developed with convergence guarantees, validated on the LIAR2 dataset \citep{xu2024enhanced} using large language models for sentiment extraction.

Our analysis demonstrates up to 76\% misinformation reduction across network configurations including radicalized environments with stubborn extremist users. Critically, we reveal engagement-misinformation trade-offs: while mean engagement may decrease, median engagement in radical networks \emph{improves}, indicating enhanced discourse quality for non-extremist majorities. The optimal operating region 
provides actionable guidance for platform operators balancing content moderation with business objectives.

\section{Methods}
\label{sec:methods}

We present a control framework for misinformation mitigation through recommender systems. The model dynamics and control-loop formulation (Section~\ref{subsec:model_dynamics}) follow \citet{sprenger2024control}; we refer readers there for foundational details. Section~\ref{subsec:cost_function} introduces our modified cost function incorporating psychological factors associated with misinformation spread. Section~\ref{subsec:mf_mb} formulates model-free and model-based control strategies. Section~\ref{subsec:convergence} provides convergence proofs.

\subsection{Model Dynamics and Control Formulation}
\label{subsec:model_dynamics}

We adopt the closed-loop Friedkin-Johnsen framework from \citet{sprenger2024control}, representing users as nodes, with overall system state $\mathbf{x}(t) \in [0,1]^n$,  at discrete time $t$, whose $i-th$ entry represents the opinion state of the $i$-th node (user). The recommender acts as an additional node influencing users through control input $u(t) \in [0,1]$.
While \citet{sprenger2024control} model topical opinion agreement, we employ the same structure to model sentiment propagation, motivated by evidence that emotional content drives misinformation virality \citep{brady2017emotion, vosoughi2018spread}. In our formulation, $x_i(t) \in [0,1]$ represents user $i$'s sentiment intensity, where $x_i(t) = 0$ corresponds to neutral/positive sentiment and $x_i(t) = 1$ to highly emotional content. The control input $u(t)$ represents recommended content sentiment.

The network is represented by row-substochastic adjacency matrix $\mathbf{W}_{\text{total}} \in [0,1]^{(n+1) \times (n+1)}$, partitioned into user-to-user interactions $\mathbf{W} \in [0,1]^{n \times n}$ and recommender-to-user influence $\mathbf{w}_{\text{rec}} \in [0,1]^n$. Dynamics evolve as:
\begin{equation}
\mathbf{x}(t+1) = (\mathbf{I}_n-\mathbf{\Lambda})\mathbf{W}\mathbf{x}(t)+(\mathbf{I}_n-\mathbf{\Lambda})\mathbf{w}_{\text{rec}}u(t)+\mathbf{\Lambda}\mathbf{x}(0),
\label{eq:dynamics1}
\end{equation}
where $\mathbf{\Lambda} = \text{diag}(\lambda_1, \ldots, \lambda_n) \in [0,1]^{n \times n}$ is the stubbornness matrix with $\lambda_i \in [0,1]$ quantifying user $i$'s resistance to influence ($\lambda_i = 0$: full susceptibility; $\lambda_i = 1$: complete adherence to $x_i(0)$). 
In compact form, 
\begin{equation}
\mathbf{x}(t + 1) = \mathbf{A}\mathbf{x}(t) + \mathbf{B}u(t) + \mathbf{\Lambda}\mathbf{x}(0),
\label{eq:dynamics2}
\end{equation}
where $\mathbf{A} = (\mathbf{I}_n - \mathbf{\Lambda})\mathbf{W}$ and $\mathbf{B} = (\mathbf{I}_n - \mathbf{\Lambda})\mathbf{w}_{\text{rec}}$ 
This linear time-invariant structure enables optimal control design for $u(t)$ balancing engagement with misinformation mitigation by penalizing high sentiment intensity.


\subsection{Cost Function Modification for Misinformation Mitigation}
\label{subsec:cost_function}

We modify the engagement objective from \citet{sprenger2024control} to incorporate misinformation mitigation while acknowledging that platforms fundamentally rely on engagement. The original engagement cost the recommender system in \cite{sprenger2024control} minimizes, 
\begin{equation}
    \theta(\mathbf{x}(t), u(t)) = \|\mathbf{x}(t) - u(t)\mathbf{1}_n\|_2^2,
    \label{eq:cost_original}
\end{equation}  
measures squared Euclidean distance between user states and recommendations and, based on confirmation bias, promotes recommnended content $u(t)$ that more closely aligns with users'opinions. We augment this with penalties for extreme sentiment intensity $E(u(t)) = \|u(t)\|^2$ and novelty modulation $N(t, t_c) = e^{-\delta(t-t_c)}$ for $t - t_c \leq z$, where $\delta > 0$ controls decay rate, $t_c$ is content creation time, and $z$ defines the content window. Novel content, which misinformation exploits \citep{berger2012makes}, receives higher initial penalty that diminishes over time. The modified cost accounting for misinformation mitigation is:
\begin{equation}
\theta_{\rm M}(\mathbf{x}(t), u(t)) = \theta(\mathbf{x}(t), u(t)) + \rho n \cdot \|u(t)\|^2 \cdot e^{-\lambda(t-t_c)}, \quad t - t_c \leq z,
\label{eq:cost_modified}
\end{equation}
where $\rho \geq 0$ controls penalty strength and $n$ ensures consistent scaling across network sizes. The recommender minimizes $\sum_{t=0}^{\infty} \theta_{\rm M}(\mathbf{x}(t), u)$; we develop tractable approximations below.

\subsection{Model-free and Model-based Approaches}
\label{subsec:mf_mb}

\textbf{Model-free (MF).} The MF approach $u_{\text{MF}}(t) = \arg\min_{u\in[0,1]} \theta_{\rm M}(\mathbf{x}(t), u)$ minimizes $\theta_{\rm M}$ at time $t$ using only $\mathbf{x}(t)$.

\noindent\textbf{Model-based (MB).} The MB approach employs model predictive control (MPC). The theoretical optimal steady-state $(\mathbf{x}^*_{\text{MB}}, u^*_{\text{MB}}) = \arg\min_{\mathbf{x},u} \theta_{\rm M}(\mathbf{x}, u)$ subject to $\mathbf{x} = \mathbf{A}\mathbf{x} + \mathbf{B}u + \mathbf{\Lambda}\mathbf{x}(0)$ and $u \in [0, 1]$ always exists (Section~\ref{subsec:convergence}). The full MPC formulation is:
\begin{align}
\mathcal{O}^*_t & := \min_{\mathbf{x}_{\xi|t}, u_{\xi|t}} \sum_{k=0}^{T-1} \theta_{\rm M}(\mathbf{x}_{k|t}, u_{k|t}) \label{eq:mpc} \\
\text{s.t. } &
\mathbf{x}_{k+1|t} = \mathbf{A}\mathbf{x}_{k|t} + \mathbf{B}u_{k|t} + \mathbf{\Lambda}\mathbf{x}(0), \; \mathbf{x}_{0|t} = \mathbf{x}(t), \; \mathbf{x}_{T|t} = \mathbf{x}_{\text{MB}}, \; u_{k|t} \in [0, 1], \; \forall k \in [0, T - 1] \nonumber
\end{align}
where $\mathcal{O}^*_t$ is the general optimization cost function and $T$ is the prediction horizon. The optimizer output $u_{\text{MB}}(t) = u_{0|t}$ is the first element of the MPC solution. The MF and MB approaches have significant differences and scopes of informational access. Unlike the MF approach, the MB approach must have access to the opinion dynamic dependencies $\mathbf{A}$, $\mathbf{B}$ and the resilience matrix $\mathbf{\Lambda}$, which is hardly measurable.

\subsection{Mathematical Analysis and Convergence Proofs}
The new cost function definition in \eqref{eq:cost_modified} requires updated analysis of the control strategies and their convergence properties. While the Friedkin-Johnsen model structure and graph properties from \citet{sprenger2024control} remain unchanged, the modified optimization problem necessitates new theoretical results. In addition, all steady-state convergence values can be seen as a region of convergence rather than a single point, this is due to the reliance of all steady-state solutions on $t - t_c$.

\subsubsection{Convergence Analysis and Steady-State Solutions}
\label{subsec:convergence}
We now derive the steady-state solutions for both the model-free and model-based approaches and establish convergence guarantees. The proofs closely follow \citet{sprenger2024control}, with modifications to accommodate the penalty terms in $\theta_{\rm M}$.

\textbf{Model-Free Steady State.} 
The optimal MF control is obtained by minimizing $\theta_{\rm M}(\mathbf{x}(t), u(t))$ with respect to $u$ at each time step. Given $\theta_{\rm M}$ is convex, taking $\frac{\partial\theta_{\rm M}}{\partial u} = 0$ and solving yields:
\begin{equation}
u^*_{\text{MF}}(t) = \frac{\sum_{i=1}^n x_i(t)}{n(1 + \rho \cdot e^{-\lambda(t-t_c)})}.
\label{eq:mf_steady_state}
\end{equation}
Substituting this into \eqref{eq:dynamics2} gives the closed-loop dynamics:
\begin{equation}
\mathbf{x}(t + 1) = (\mathbf{I}_n - \mathbf{\Lambda})\mathbf{F}\mathbf{x}(t) + \mathbf{\Lambda}\mathbf{x}(0),
\label{eq:mf_dynamics}
\end{equation}
where $\mathbf{F} = \mathbf{W} + \frac{\mathbf{w}_{\text{rec}}\mathbf{1}_n^T}{n(1 + \rho \cdot e^{-\lambda(t-t_c)})}$. The matrix $\mathbf{F}$ is sub-row stochastic and satisfies the convergence conditions established in \citet{sprenger2024control}. At steady state, $\mathbf{x}(t) = \mathbf{x}(t+1)$, which yields:
\begin{equation}
\mathbf{x}^*_{\text{MF}} = \left(\mathbf{I}_n - \mathbf{A} - \frac{\mathbf{B} \cdot \mathbf{1}_n^T}{n(1 + \rho \cdot e^{-\lambda(t-t_c)})}\right)^{-1} \mathbf{\Lambda}\mathbf{x}(0).
\label{eq:mf_user_steady_state}
\end{equation}

\textbf{Model-Based Steady State.} 
For the MB approach, we solve the constrained optimization problem \eqref{eq:mpc} using the Karush-Kuhn-Tucker conditions. The interior solution ($0 < u < 1$) is:
\begin{equation}
u^*_{\text{MB}} = \frac{\mathbf{1}_n^T \mathbf{y} - \mathbf{v}^T \mathbf{y}}{-\mathbf{1}_n^T \mathbf{v} + n + \mathbf{v}^T \mathbf{v} - \mathbf{v}^T \mathbf{1}_n + \rho n e^{-\lambda(t-t_c)}}, \quad
\mathbf{x}^*_{\text{MB}} = \mathbf{v}u^*_{\text{MB}} + \mathbf{y},
\label{eq:mb_steady_states}
\end{equation}
where $\mathbf{v} = (\mathbf{I}_n - \mathbf{A})^{-1}\mathbf{B}$ and $\mathbf{y} = (\mathbf{I}_n - \mathbf{A})^{-1}\mathbf{\Lambda}\mathbf{x}(0)$.

\textbf{Convergence Guarantees.} 
The convergence proof for both approaches follows the same structure as \citet{sprenger2024control}. The key modification is in the matrix:
\begin{equation}
\mathbf{H} = \begin{bmatrix}
\mathbf{I}_n & -\mathbf{1}_n \\
-\mathbf{1}_n^T & n(1 + \rho e^{-\lambda(t-t_c)})
\end{bmatrix},
\label{eq:H_matrix}
\end{equation}
which replaces their corresponding matrix in the Lyapunov stability analysis. All other proof steps remain identical, and we refer readers to \citet{sprenger2024control} for the complete argument. The penalty terms $\rho$ and $e^{-\lambda(t-t_c)}$ preserve the positive definiteness of $\mathbf{H}$ required for convergence, provided $\rho \geq 0$ and $\lambda > 0$.

\section{Simulation Setup}
\label{sec:simulation}

This section details the simulation design, network configurations, and datasets used to evaluate the proposed misinformation mitigation framework. Two types of networks are considered: a large-scale synthetic network of 100 agents, and a small radicalized network of 6 agents adapted from \citet{sprenger2024control}. Each network is tested using both synthetic continuous content and real-world data from the LIAR2 dataset.

\subsection{Network Configurations}
\label{subsec:network_configs}

\textbf{Network A — 100-agent synthetic network.}  
This network models a general social platform with $n=100$ users. Network parameters are given in Table~\ref{tab:simulation_params}. The initial opinion (sentiment) values are drawn from a beta distribution of parameters $\alpha=7, \beta=2$, which is skewed toward higher sentiment values (more emotional intensity), reflecting populations where mildly negative content dominates engagement. This setup is used to evaluate overall mitigation performance.

\noindent \textbf{Network B — 6-agent radicalized network.}  
Following the structure of \citet{sprenger2024control}, we consider a smaller network of six users to study the influence of a stubborn radical agent. The initial opinions are defined as the complement of those used in the original paper, so that the most stubborn user now holds an extreme negative opinion of $1$.

Specifically, $\mathbf{x}(0) = [0.33, 0.26, 0.17, 0.32, 1.00, 0.41]$. This modification ensures that the controller faces the more challenging task of mitigating an entrenched source of negativity. Unlike the large-scale network, this setup runs for $\tau=50$ time steps, which is sufficient for convergence.

\begin{table}[t]
\centering
\caption{Simulation Parameters for Both Network Configurations}
\label{tab:simulation_params}
\begin{tabular}{lll}
\hline
\textbf{Parameter} & \textbf{Description} & \textbf{Value} \\
\hline
$n$ & Number of users & 100 (Network A) / 6 (Network B) \\
$\Lambda_h$ & Highest stubbornness & 0.05 \\
$\Lambda_l$ & Lowest stubbornness & 0.00 \\
$\kappa_u$ & User-to-user connectivity & 0.25 \\
$\kappa_r$ & Recommender-to-user connectivity & 0.80 \\
$\tau$ & Time steps & 100 (A) / 50 (B) \\
$\rho$ & Penalty strength regulator & [0.00, 5.50] (step 0.10) \\
$T$ & MPC prediction horizon & 50 \\
$z$ & Eligible content window & 5 \\
$\lambda$ & Novelty decay rate & 0.00 \\
\hline
\end{tabular}
\end{table}

\subsection{Simulation Scenarios}
\label{subsec:scenarios}

Each network is simulated under three configurations:
\begin{enumerate}
    \item \textbf{Model-Free (MF)} without mitigation ($\rho=0$), optimizing only user engagement $\theta$, as in \cite{sprenger2024control}.
    \item \textbf{Model-Free (MF)} with mitigation ($\rho>0$), introducing the misinformation penalty,  $\theta_{\rm M}$.
    \item \textbf{Model-Based (MB)} with mitigation ($\rho>0$), using the predictive control formulation.
\end{enumerate}
For all cases, we consider both synthetic continuous $u(t)\in[0,1]$ and discrete data-driven content values described below.

\subsection{Data-based Sentiment Extraction}
\label{subsec:llm_data}

To simulate realistic content, we use the LIAR2 dataset \citep{xu2024enhanced}, which contains 4000 labeled statements (2000 true and 2000 false) from social media and news sources. For each statement, we compute a \emph{emotional extremity score} $C(l)\in[0,1]$ using a transformer-based natural language processing (NLP) model (Mistral NeMo). The model analyzes each text along six affective and linguistic dimensions—fear, disgust, anxiety, shock, overall negative sentiment, and subjectivity—and aggregates them into a single weighted score by giving equal weight (0.15) to each emotional dimension and slightly higher weight (0.20) to overall negativity and subjectivity. Higher $C(l)$ values correspond to content with stronger emotional tone or higher potential for emotional manipulation. The Mistral NeMo model is chosen for its efficiency, accuracy, and open-source availability.
At each simulation time step, a random subset of the 4000 LIAR2 statements is made available to the recommender. On average, this corresponds to about 40 new pieces of content per step in the 100-agent network and about 80 in the 6-agent network. The appearance times are uniformly distributed to emulate a continuous stream of new posts. Consistent with prior research on misinformation virality, false statements exhibit higher average emotional intensity (mean $C(l) = 0.537$) compared to true statements (mean $C(l) = 0.379$).

\subsection{Misinformation and Behavioral Metrics}
\label{subsec:metric}

To evaluate the effectiveness of the proposed mitigation strategies, we monitor three complementary quantities.  $i)$
{First}, we compute the \emph{misinformation metric}
\begin{equation}
\mathcal{M} = \frac{\# {\rm false} {\rm news}}{\# \rm{news}},
\label{eq:mitigation_metric}
\end{equation}
as the ration of false news. A lower value of $\mathcal{M}$ indicates stronger suppression of misinformation exposure. 
$ii)$ {Second}, we quantify the overall \emph{sentiment shift} as the absolute change in emotional extremity between the final and initial states, i.e.\ the mean (and median) of $|x_i(\tau) - x_i(0)|$ across users. This captures how much individual sentiment evolves during the simulation. \\
$iii)$ {Finally}, we track user \emph{engagement}, defined as the per-user average of the instantaneous engagement cost introduced in \eqref{eq:cost_original}, averaged over time. Higher engagement values correspond to stronger alignment between user sentiment and recommended content. 

\section{Results}
\label{sec:results}

\subsection{Misinformation Mitigation Alleviates Users'Emotional Extremity}

\begin{figure}[t]
    \centering
    \includegraphics[scale=0.3]{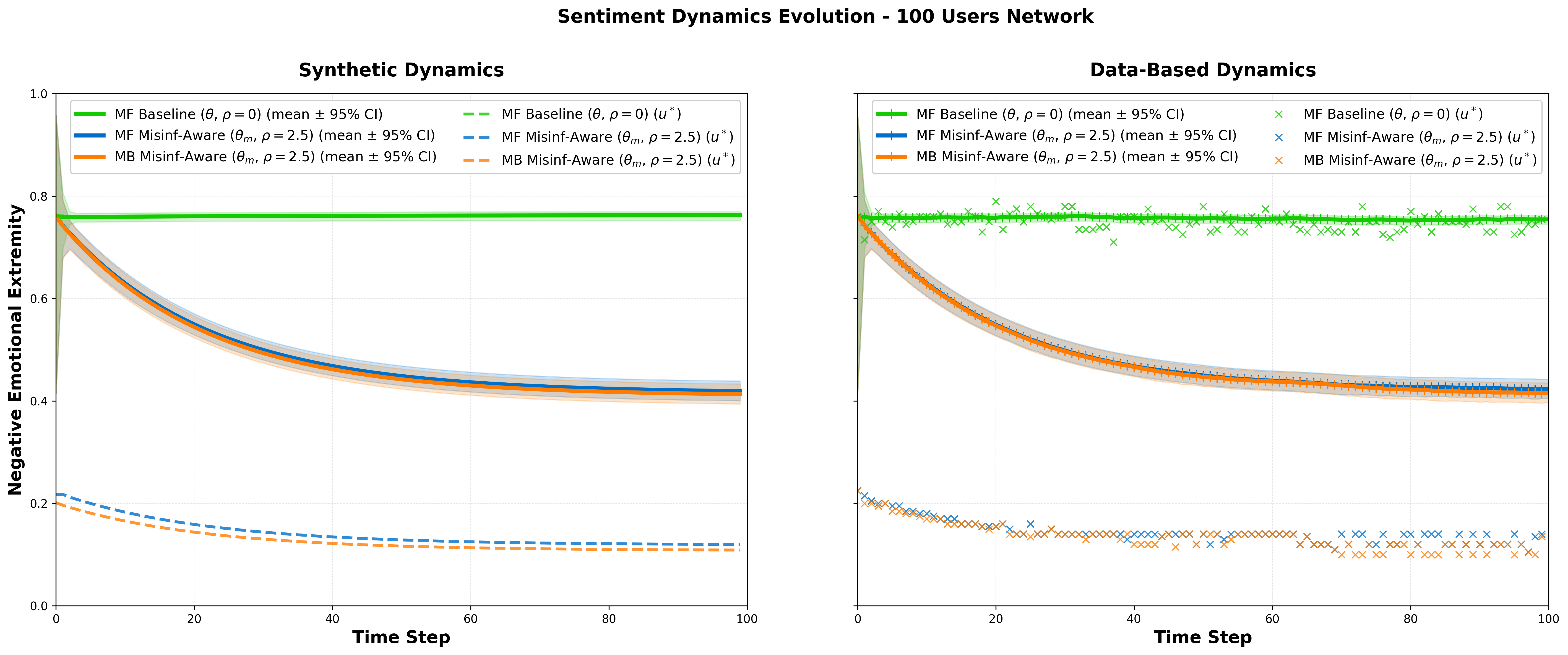}
       \includegraphics[scale=0.3]{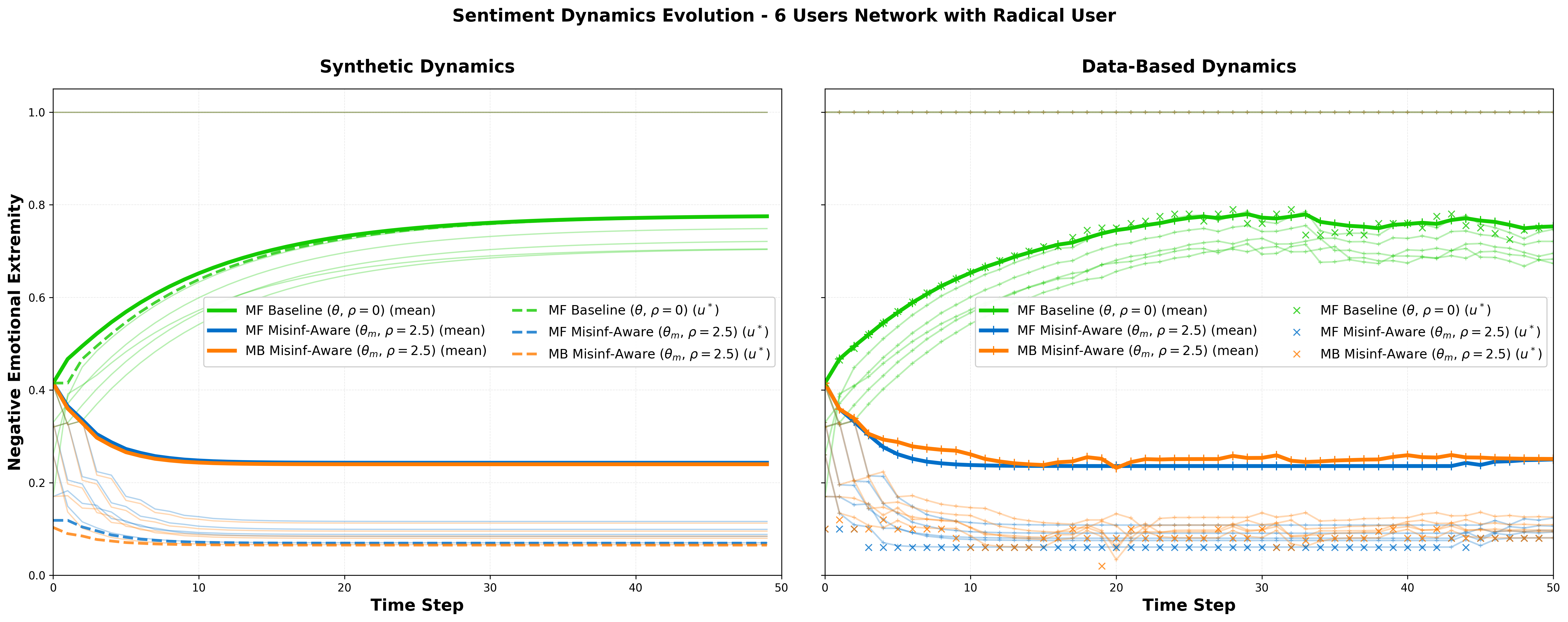}
    \caption{Sentiment Dynamics Evolution. \textbf{Top}: 100-User Network showing mean user emotional extremity (solid) and recommender output (dashed) over 100 time steps, comparing baseline engagement-only control ($\theta$, green) with MF mitigation ($\theta_{\rm M}$, blue) and MB mitigation ($\theta_{\rm M}$, orange) at $\rho=2.5$. Left: synthetic continuous dynamics; Right: data-driven discrete content. Shaded regions indicate standard deviation. \textbf{Bottom}: 6-User Network with Radical User showing the same comparison over 50 time steps. Individual user trajectories are shown in light lines. The stubborn radical user remains at maximum emotional extremity (top of plot), while mitigation strategies prevent negativity propagation to other users.}
    \label{fig:opinion_dynamics}
\end{figure}

Figure~\ref{fig:opinion_dynamics} shows the evolution of user emotional extremity under different control strategies for both network configurations. In both networks, the baseline engagement-only model ($\theta$) drives average sentiment toward more negative values, while the mitigation-aware controllers ($\theta_{\rm M}$) stabilize user states closer to neutrality.
{In the 100-agent network} (Figure~\ref{fig:opinion_dynamics}, top), the average emotional extremity converges rapidly to a steady moderate value. The MF and MB mitigation strategies exhibit nearly identical behavior, both successfully preventing the negative drift observed in the baseline case. This holds true for both synthetic continuous dynamics (left panels) and data-driven discrete content selection (right panels), demonstrating robustness across simulation modalities.
{In the 6-agent network} (Figure~\ref{fig:opinion_dynamics}, bottom), the presence of a stubborn radical user anchored at $x_i = 1$ (maximum emotional extremity) causes the overall mean sentiment to remain relatively high, yet still substantially less extreme than without mitigation. Notably, the mitigation controllers prevent the radical user's negativity from propagating to connected users, maintaining their sentiment at moderate levels despite the persistent influence of the extremist node.

\subsection{Effect of trade-off Parameter $\rho$ on Misinformation Spreading}\label{sec:tradeoff}

\begin{figure}[t]
    \centering
    \includegraphics[width=\linewidth]{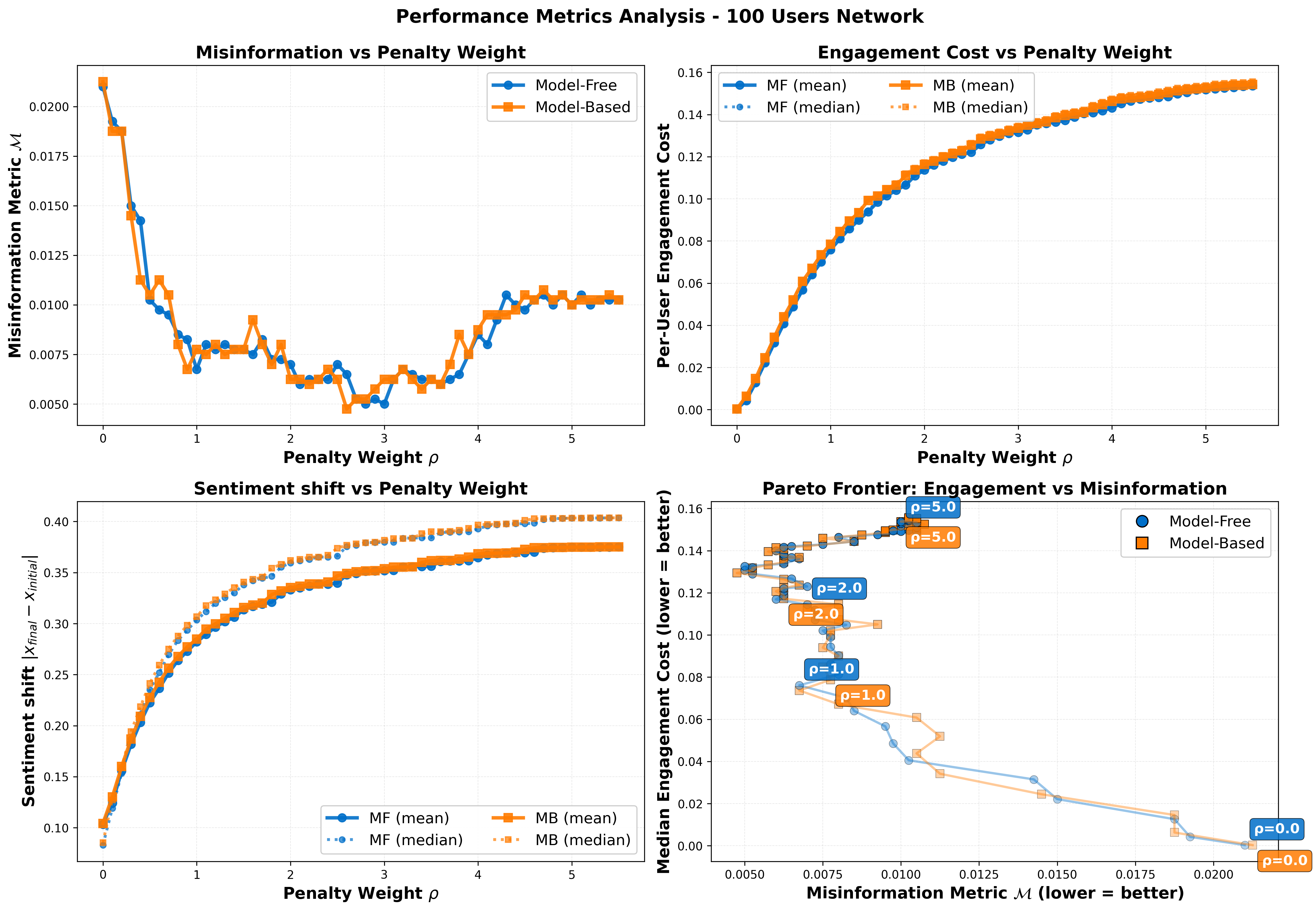}
    \caption{Performance Metrics Analysis for 100-User Network (Data-Driven). \textbf{Top-left}: Misinformation metric $\mathcal{M}$ vs.\ penalty weight $\rho$. \textbf{Top-right}: Per-user engagement cost vs.\ $\rho$ (mean and median). \textbf{Bottom-left}: Sentiment shift $|x_i(\tau) - x_i(0)|$ vs.\ $\rho$ (mean and median). \textbf{Bottom-right}: Pareto frontier showing trade-off between median per-user engagement cost and misinformation (lower-left is better). Labeled points indicate $\rho$ values. Blue: Model-Free; Orange: Model-Based.}
    \label{fig:metrics_standard}
\end{figure}

Figure~\ref{fig:metrics_standard} illustrates how varying the penalty coefficient $\rho$ influences misinformation spread and system behavior in the 100-user network under data-driven content selection.
The mitigation metric $\mathcal{M}$ (top-left) decreases monotonically as $\rho$ increases from 0 to approximately 2.5, corresponding to a $\sim$76\% reduction in misinformation spread compared to baseline ($\rho=0$). Beyond $\rho > 3$, performance slightly degrades, likely due to LLM misclassification of emotionally neutral yet false statements (discussed in Section~\ref{sec:discussion}).
Per-user engagement cost (top-right) increases with $\rho$ for both mean and median, but plateaus around $\rho \approx 3$, indicating users maintain substantial alignment with recommendations despite prioritization of less emotionally extreme content.
Sentiment shift (bottom-left) increases with $\rho$, reflecting stronger moderation of initial emotional extremity. MF and MB approaches yield similar trajectories, with MB showing marginally higher shift at large $\rho$ due to predictive optimization.
The Pareto frontier (bottom-right) reveals that substantial misinformation reduction is achievable with modest engagement cost increases. The optimal operating region $\rho \in [1.0, 2.5]$ balances both objectives. MF and MB approaches trace nearly identical curves, indicating the simpler MF strategy suffices.

\begin{figure}[t]
    \centering
    \includegraphics[width=\linewidth]{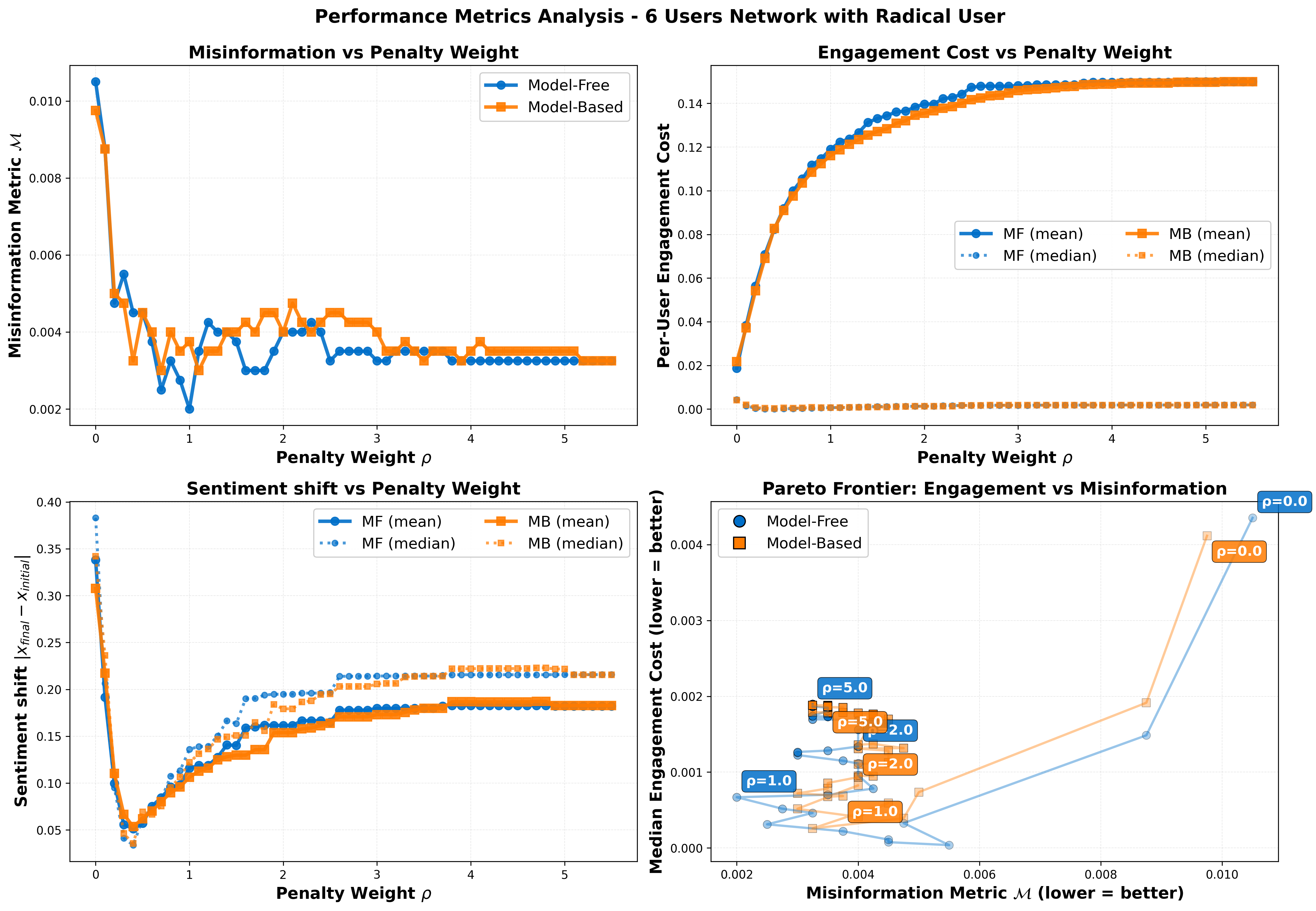}
    \caption{Performance Metrics Analysis for 6-User Radical Network (Data-Driven). \textbf{Top-left}: Misinformation metric $\mathcal{M}$ vs.\ $\rho$. \textbf{Top-right}: Per-user engagement cost vs.\ $\rho$; note that median (dashed) remains stable while mean (solid) increases, indicating improved engagement for non-radical users. \textbf{Bottom-left}: Sentiment shift vs.\ $\rho$. \textbf{Bottom-right}: Pareto frontier using median engagement, showing that mitigation can improve both objectives simultaneously in radicalized networks. Blue: Model-Free; Orange: Model-Based.}
    \label{fig:metrics_radical}
\end{figure}

\subsection{A Special Focus on Radical Users}

Figure~\ref{fig:metrics_radical} presents the same analysis as in Section~\ref{sec:tradeoff} for the 6-agent radicalized network, revealing important differences from the large network case.
{Misinformation mitigation} (top-left) follows a similar trend, with optimal performance around $\rho = 1$, achieving up to {$\sim$70\% reduction} compared to the baseline that only accounts for engagement maximization. 
{Engagement dynamics} (top-right): Unlike the large network, {median engagement remains nearly constant} across all $\rho$ values (dashed orange line), while {mean engagement increases} (solid blue line). This divergence suggests that while the radical user's engagement decreases (pulling up the mean) with increasing $\rho$, the majority of users engage \emph{more} with moderated content.
{Sentiment shift} (bottom-left): Compared to the 100-user case, interestingly the sentiment shifts (mean and median) are not monotonically increasing with $\rho$, rather a minimum is achieved approximately when $\rho=0.4$.
{Pareto frontier} (bottom-right): In this case, when considering median engagement instead of mean, the trade-off curve inverts: higher $\rho$ values simultaneously reduce misinformation \emph{and} maintain or improve median engagement. This indicates that for networks with extremist users, mitigation strategies can enhance discourse quality for the majority of participants.

\section{Discussion}
\label{sec:discussion}

Penalizing emotionally extreme recommendations through $\theta_{\rm M}$ effectively reduces misinformation while maintaining engagement. Both MF and MB controllers converge to predicted equilibria, demonstrating robustness across network sizes. Their nearly identical performance suggests the simpler MF strategy suffices for deployment.
The optimal penalty $\rho \in [1.0, 2.5]$ achieves up to 76\% misinformation reduction, providing practical guidance for platforms balancing moderation with engagement.
The radical network case reveals key insights: while misinformation control cannot override stubborn extremists, negativity propagation to other users is significantly reduced. Moreover, median engagement \emph{improves} under mitigation, suggesting content moderation enhances discourse quality for non-radical majorities in polarized environments.
Performance degradation at $\rho > 3$ stems from linguistically neutral false statements in LIAR2, weakening the emotion-truthfulness correlation. This emphasizes the need for datasets with fine-grained truth levels, temporal dynamics, and diverse linguistic styles targeting boundary cases where misinformation employs objective framing.
Future work should integrate $\rho$ and $\mathcal{M}$ into the closed-loop for adaptive control responding to misinformation surges during elections or crises. Incorporating time-dependent novelty factors with temporal shareability data could improve viral content responsiveness. Specialized LLMs trained on misinformation corpora could enhance classification. Finally, field experiments would validate theoretical predictions and reveal practical implementation challenges.

\section{Conclusions}
\label{sec:conclusions}

This paper presents a control framework for mitigating misinformation through sentiment-aware recommender systems. By adapting Friedkin-Johnsen dynamics to represent emotional extremity and penalizing characteristics misinformation exploits, we demonstrate up to 76\% reductions in misinformation spread while maintaining engagement.
Key contributions include: (1) a modified cost function $\theta_{\rm M}$ penalizing misinformation-associated content characteristics; (2) convergence guarantees for both model-free and model-based strategies; (3) validation using LIAR2 dataset with LLM-extracted sentiment features; and (4) evidence that content moderation improves median engagement for non-extremists in radicalized networks.
The framework provides foundation for next-generation recommender systems accounting for emotional and cognitive propagation dynamics. While challenges remain, e.g., dataset quality, adaptive tuning, results suggest algorithmic interventions can address misinformation without abandoning engagement-driven models. Future work should focus on real-world deployment, adaptive mechanisms, and integration with complementary strategies like fact-checking and user education.

\section*{Data and Code Availability}

The code implementing the proposed framework and the processed LIAR2 dataset with extracted sentiment features are publicly available at \href{https://github.com/paganick/misinformation-mitigation-model}{https://github.com/paganick/misinformation-mitigation-model}.

\bibliography{references}

\end{document}